\documentclass[prb,twocolumn,showpacs,amsmath,amssymb,superscriptaddress,floatfix,intlimits]{revtex4-1}

\usepackage{graphicx}
\usepackage{braket}
\usepackage{mathtools}
\usepackage[normalem]{ulem}
\usepackage{color,soul}
\usepackage{hyperref}
\hypersetup{pdfnewwindow=true, colorlinks=true, linkcolor=blue, anchorcolor=blue, citecolor=blue, filecolor=blue, menucolor=blue, urlcolor=blue}
\usepackage{amsmath}
\usepackage{physics}
\usepackage{siunitx}

\newcommand{\GoWo}{G$_0$W$_{0}$}

\begin{document}

\title{Comparison of GW band structure to semi-empirical approach for an FeSe monolayer}

\author{Diana Y. Qiu}
\affiliation{Department of Physics, University of California at Berkeley, California 94720, USA
and Materials Sciences Division, Lawrence Berkeley National Laboratory, Berkeley, California 94720, USA}
\author{Sinisa Coh}
\affiliation{Materials Science and Engineering and Mechanical Engineering, University of California Riverside, Riverside, CA 92521, USA}
\author{Marvin L. Cohen}
\affiliation{Department of Physics, University of California at Berkeley, California 94720, USA
and Materials Sciences Division, Lawrence Berkeley National Laboratory, Berkeley, California 94720, USA}
\author{Steven G. Louie}
\affiliation{Department of Physics, University of California at Berkeley, California 94720, USA
and Materials Sciences Division, Lawrence Berkeley National Laboratory, Berkeley, California 94720, USA}

\date{\today}

\begin{abstract}
We present the \GoWo{} band structure, core levels, and deformation potential of monolayer FeSe in the paramagnetic phase based on a starting mean field of the Kohn Sham density functional theory (DFT) with the PBE functional. We find the GW correction increases the bandwidth of the states forming the $M$ pocket near the Fermi energy, while leaving the $\Gamma$ pocket roughly unchanged. We then compare the \GoWo{} quasiparticle band energies with the band structure from a simple empirical +A approach, which was recently proposed to capture the renormalization of the electron-phonon interaction going beyond DFT in FeSe, when used as a starting point in density functional perturbation theory (DFPT). We show that this empirical correction succeeds in approximating the GW non-local and dynamical self energy in monolayer FeSe and reproduces the GW band structure near the Fermi surface, the core energy levels, and the deformation potential (electron-phonon coupling).
\end{abstract}

\maketitle

\section{Introduction}

The report of superconducting transition temperatures ($T_{\rm c}$) as high as 100~K in monolayer FeSe on SrTiO$_3$ (STO) has inspired a wide range of interest in understanding its electronic properties and the origin of the high $T_{\rm c}$~\cite{wang12,liu12,zhang14,deng14,liu14,ge15,tian2016}. These high $T_{\rm c}$ are notable for being much higher than that of bulk FeSe ($T_{\rm c}$=8~K)\cite{hsu2008} and other Fe-based superconductors, such as SmO$_x$F$_{1-x}$FeAs ($T_{\rm c}$=55~K)~\cite{ren2008} and A$_x$Fe$_{2-y}$Se$_2$ ($T_{\rm c}=$30 K)\cite{mou2011}. Angle-resolved photoemission spectroscopy (ARPES) reveals that doped monolayer FeSe supported on STO and other oxide subbstrates has a Fermi surface consisting only of very small electron pockets at the corners of the Brillouin zone (the M point)~\cite{he13,liu14,lee14}, distinct from both bulk FeSe and bilayer FeSe, which both possess an additional hole pocket around the $\Gamma$ point~\cite{lee14,lee2015}. Researchers have attempted to understand the high $T_{\rm c}$ in supported monolayer FeSe through a combination of explanations involving charge transfer from the substrate ~\cite{he13,miyata2015,shiogai2015,lei2016,wen2016,lei2016b} and coupling to interfacial phonon modes~\cite{lee14,rebec2017,rademaker2016,tian2016}.

Unfortunately, understanding the electronic structure of FeSe is complicated by the fact that standard first-principles approaches, like the semi-local generalized gradient approximation (GGA) to the exchange within density functional theory (DFT), give results that do not agree with experimental measurements of electronic~\cite{chen2010,tomai2010,maletz2014,nakayama2010}, structural~\cite{subedi2008}, or magnetic properties~\cite{subedi2008,li2009} for Fe-based superconductors. For the electronic properties, it is well-known that DFT overestimates the bandwidth of the $M$-point electron pocket in FeSe compared to experiment. This overestimation of the bandwidth is a problem common to DFT calculations on metallic systems and can be corrected by accounting for electron-electron interactions in the self energy at higher levels of theory, such as GW~\cite{northrup89}. GW and dynamical mean field theory (DMFT) calculations on bulk FeSe result in band narrowing and improved agreement with the experimental bandwidths and the magnetic ground state~\cite{aichhorn10,yin11,tomczak12}. 

Remarkably, adding a simple empirical correction to GGA at the PBE level (GGA+A) selects a ground state of FeSe that is largely consistent with experiment and greatly enhances the deformation potential, resulting in a concomitant increase in the electron-phonon coupling in DFPT calculations, in good agreement with inelastic tunneling data~\cite{coh2015}. This approximation of the self energy by a simple local potential on the Fe sites can be justified if the self energy in FeSe is mostly local in real space, as shown to be true in Refs.~\onlinecite{zein2006,tomczak12}, and largely frequency independent. Here, we evaluate the accuracy of the GGA+A approach by comparing the electronic structure of monolayer FeSe obtained within GGA+A with the electronic structure from the \textit{ab initio} \GoWo{} approach which employs a non-local and frequency-dependent self energy. We focus here on the isolated FeSe monolayer in the nonmagnetic phase, leaving consideration of the antiferromagnetic phase to future work. We address how different treatments of the frequency-dependence in the GW self energy affect the electronic structure of monolayer FeSe and find that the GW approach increases the effective mass of the electron pocket at the $M$ point by a factor of 1.5 compared to the effective mass at the GGA-PBE level and that the GW approach leaves the $\Gamma$ pocket mostly unchanged, compared to GGA-PBE. Finally, we compare our \GoWo{} results with GGA+A~\cite{coh2015} and find that the latter correction to DFT-PBE can accurately reproduce the GW band structure both for low-lying states and states near the Fermi level, suggesting that the self energy can be well-approximated by a local, static potential for the states in this material.

The paper is organized as follows. In section II, we discuss our computational methodology. In section III, we present the calculated GW band structure for monolayer FeSe. In section, IV, we present results for DFT with an empirical local correction on the Fe sites discussed above and compare with the GW results. We summarize in section V.

\section{Method}

The mean field starting point for our \textit{ab initio} \GoWo{} calculation~\cite{hybertsen86} is obtained from density functional theory (DFT)~\cite{hohenberg64,kohn65}, as implemented in Quantum ESPRESSO~\cite{espresso}, in the generalized-gradient approximation (GGA) for the exchange-correlation energy functional as proposed by Perdew, Burke and Ernzerhof (PBE) ~\cite{perdew96}.  The in-plane lattice constant is fixed to the lattice constant of SrTiO$_3$, and the atomic positions are fully relaxed. The calculation uses a supercell geometry, optimized norm-conserving Vanderbilt (ONCV) pseudopotentials from the library of D.R.~Hamann~\cite{hamann13} with the $3s$ and $3p$ semicore states included as valence states of the Fe atom, and a wavefunction cutoff of 100~Ry. The FeSe monolayer is doped with 0.24~electrons per unit cell to represent the doping of FeSe on STO. The dimension of the supercell is 15~$\AA$ in the out-of-plane direction.  While the experimental magnetic ground state of FeSe is under debate, all calculations here are done for the non-magnetic ground state of FeSe, so that all approaches can be compared within the same electronic ground state.

\section{\GoWo{} Band structure}

\begin{figure*}[tp]
  \includegraphics{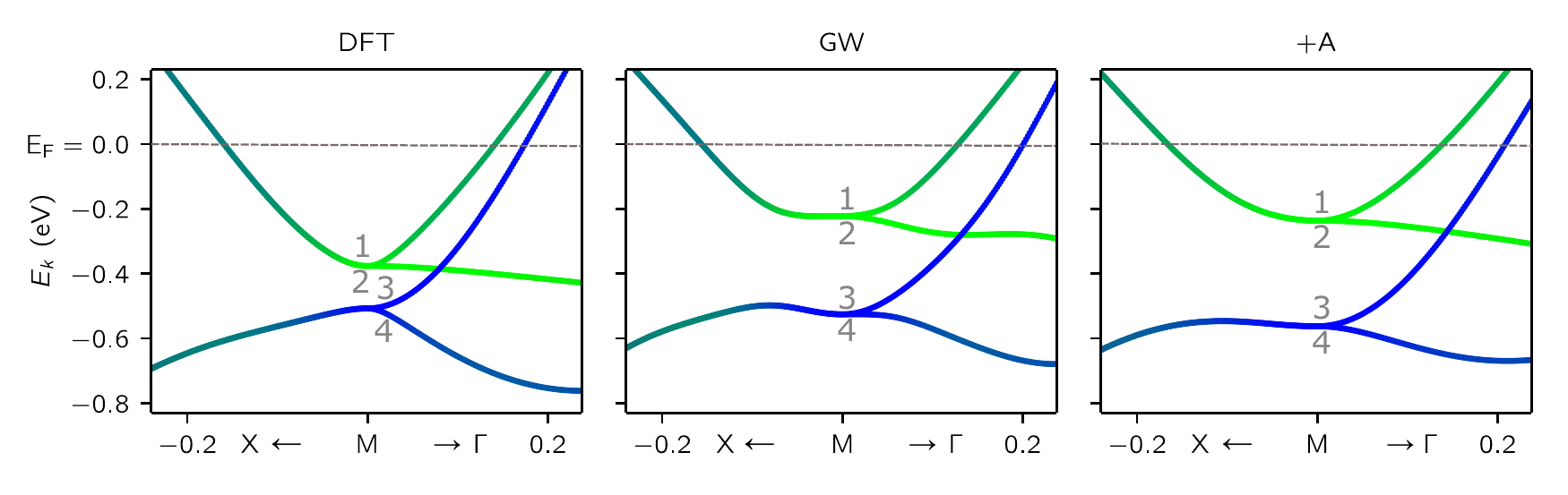}
  \caption{\label{fig:bsM} The band structure of the electron pocket near the $M$ point with path taken along the $X\rightarrow M\rightarrow \Gamma$ direction at different levels of theory.  Different colors indicate contributions from different atomic orbitals (amount of green color is proportional to the contribution of d$_{zx}$ and d$_{zy}$, blue to d$_{xy}$ and  d$_{x^2 - y^2}$, and red to the contribution from d$_{z^2}$, which is negligible). The Fermi level ($E_F$) is set at zero. The $GW$ bands are calculated at the \GoWo{} level with frequency-dependence in the dielectric screening treated within the HL-GPP model.}
\end{figure*} 

\begin{figure*}[tp]
\includegraphics{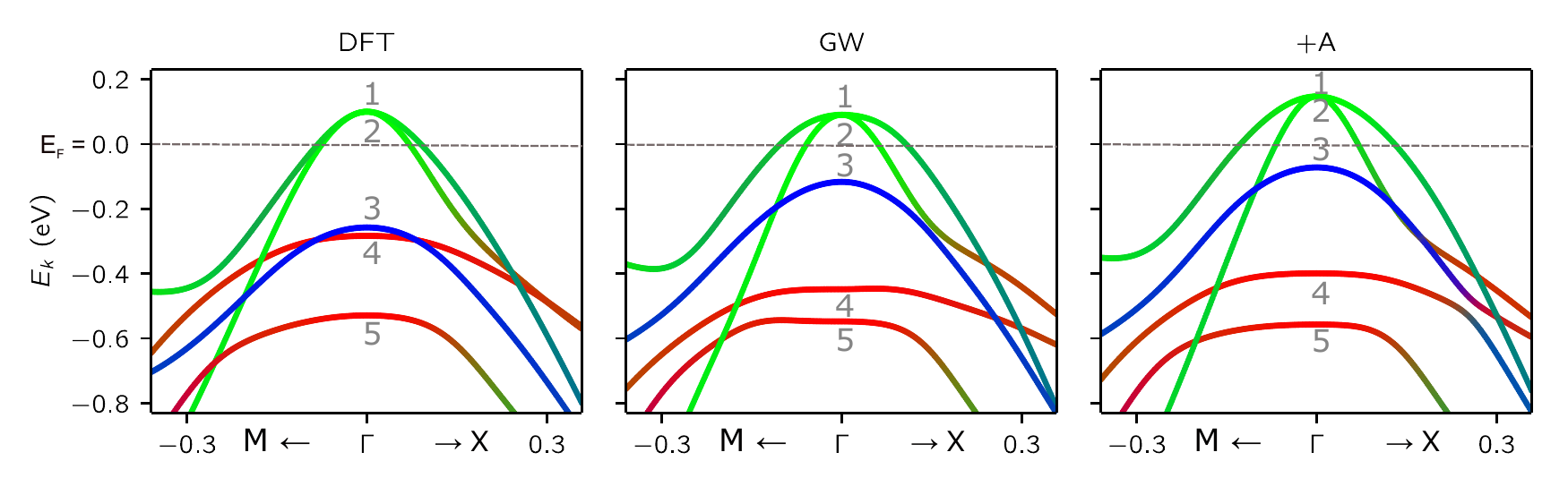}
\caption{Band structure of the hole pocket near the $\Gamma$ point with path taken along the $M\rightarrow\Gamma\rightarrow X$ direction at different levels of theory as in Fig.~\ref{fig:bsM}. Different colors indicate contributions from different atomic orbitals (the amount of green color is proportional to the contribution of d$_{zx}$ and d$_{zy}$, blue to d$_{xy}$ and  d$_{x^2 - y^2}$, and red to d$_{z^2}$). The Fermi level ($E_F$) is set at zero.}
\label{fig:bsGamma}
\end{figure*}

We first focus on the band structure, plotted using Wannier interpolation,  of monolayer FeSe near the $M$ point as shown in Fig.~\ref{fig:bsM} and near the $\Gamma$ point as shown in Fig.~\ref{fig:bsGamma}.

There are several bands that cross the Fermi surface near the M point. The inner band (labelled 1 in Fig.~\ref{fig:bsM}) consists of mostly Fe $d_{zx}$ and $d_{zy}$ character (colored green) and forms an electron pocket, which we will refer to as the $M$-pocket. The bandwidth, $E_M$, is defined as the energy difference between the bottom of the $M$ pocket and the Fermi energy $E_F$, which is set to $0$~eV. At the DFT-GGA level, $E_M=0.383$~eV. Including the self energy at the GW level reduces the occupied bandwidth $E_M$. With a one-shot \GoWo{} correction, the $M$-pocket width is reduced by 0.132~eV when the frequency dependence is approximated by the HL-GPP model~\cite{hybertsen86} and by 0.096~eV when the full frequency dependence of the dielectric screening is included in the self energy~\cite{contourdef,contourdef2}. The second highest band (labeled 2) is also composed of mainly Fe $d_{zx}$ and $d_{zy}$ character (green color) and crosses the Fermi energy along the $M$ to $X$ direction only, where it is degenerate with band 1.

\begin{figure*}[tp]
  \includegraphics{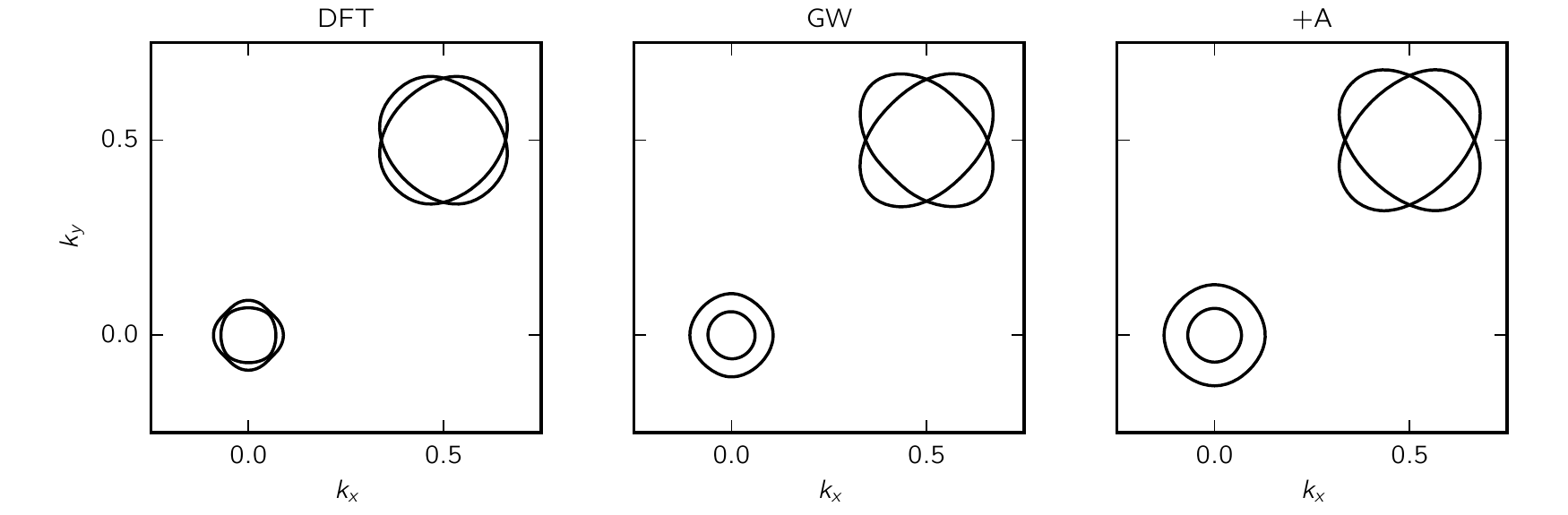}
  \caption{\label{fig:fermi} Fermi surface with pockets at $\Gamma$ and $M$ ($k_x=k_y=0.5$) points at different levels of theory. The $GW$ results are calculated at the \GoWo{} level with frequency-dependence in the dielectric screening treated within the HL-GPP model.}
\end{figure*}

Two lower bands at the $M$ point, labelled 3 and 4, consist of mainly Fe $d_{xy}$ and $d_{x^2-y^2}$ character (blue color) and lie below the bottom of the electron $M$-pocket. We label the energy difference between band 3 (or band 4, with which it is degenerate at $M$) and the bottom of the electron $M$-pocket, as $\delta_M$, and $\delta_M$ is 0.131~eV at the DFT-GGA level. The energy difference $\delta_M$ increases when the GW self-energy correction is included. However, this is quite sensitive to the treatment of the frequency-dependence of the dielectric screening in the self energy. With a generalized plasmon pole model, $\delta_M$ increases to 0.223~eV at the \GoWo{} level. When the full frequency dependence is used, however, we see that $\delta_M$ is relatively unchanged from DFT-GGA, increasing only to 0.151~eV. Band 3 crosses the Fermi surface along the $M$ to $\Gamma$ direction. At the GW level, this crossing is moved further away from $M$ toward $\Gamma$, consistent with an elongation of the Fermi surface along the $M$ to $\Gamma$ direction. The Fermi surfaces are shown in Fig.~\ref{fig:fermi}.

The band structure near the $\Gamma$ point is shown in Fig.~\ref{fig:bsGamma}. The depth of the $\Gamma$ pocket, which we label $E_\Gamma$, is roughly the same at the GGA and GW levels.  The bandwidth decreases slightly by 0.015~eV when the generalized plasmon pole~\cite{hybertsen86} is used in the GW calculation and increases slightly by 0.006~eV, when the full frequency dependence is used~\cite{contourdef,contourdef2}. Including the self-energy effects at the GW level does not eliminate the $\Gamma$ pocket, but experimentally no $\Gamma$ pocket is observed in ARPES measurements. We assign this discrepancy to the fact that our GW calculations do not include the effect of antiferromagnetic fluctuations, which remove the $\Gamma$-pocket from the calculations, even at the GGA level~\cite{Coh2016}.

The Fermi surfaces at the GGA and \GoWo{} HL-GPP levels are shown in Fig.~\ref{fig:fermi}. The Fermi surface changes considerably at the GW level, becoming larger and more elongated at the $M$ point and becoming larger with no band crossing at the $\Gamma$ point.

ARPES experiments for monolayer FeSe on STO report a Fermi surface that consists only of small electron pockets around the $M$ point, which has an occupied bandwidth, $E_M$ of 0.06--0.08~eV~\cite{liu12,he13,lee14}. Direct comparison with the experimental band structure is difficult because it is not clear whether the ground state of monolayer FeSe is antiferromagnetic or paramagnetic. ARPES spectra of monolayer FeSe on STO closely resemble the DFT band structure of the paramagnetic ground state near the $M$ point but also resemble the DFT band structure of the checkerboard antiferromagnetic ground state near the $\Gamma$ point where the band forming the hole pocket in the nonmagnetic state is pushed completely below the Fermi level~\cite{tan2013,nakayama2014,shimojima2014,watson2015,Coh2016}. There are also suggestions that the surface termination of STO may remove the $\Gamma$ pocket in the nonmagnetic state~\cite{zou16}. However, if we move the $\Gamma$-pocket below the Fermi level---in order to mimic the effect of electron transfer from STO ---and recalculate the occupied bandwidth at $M$, the bandwidth decreases to 0.2~eV, which is still about twice the experimental width. Thus, like the case of bulk FeSe~\cite{tomczak12}, the GW approximation does not capture the full renormalization of the $M$ pocket in monolayer FeSe, at least if one assumes that there is no influence by the STO substrate other than being a source of electrons and strain.

\section{Comparison of GW results with GGA+A results}

\begin{figure}[tp]
  \includegraphics{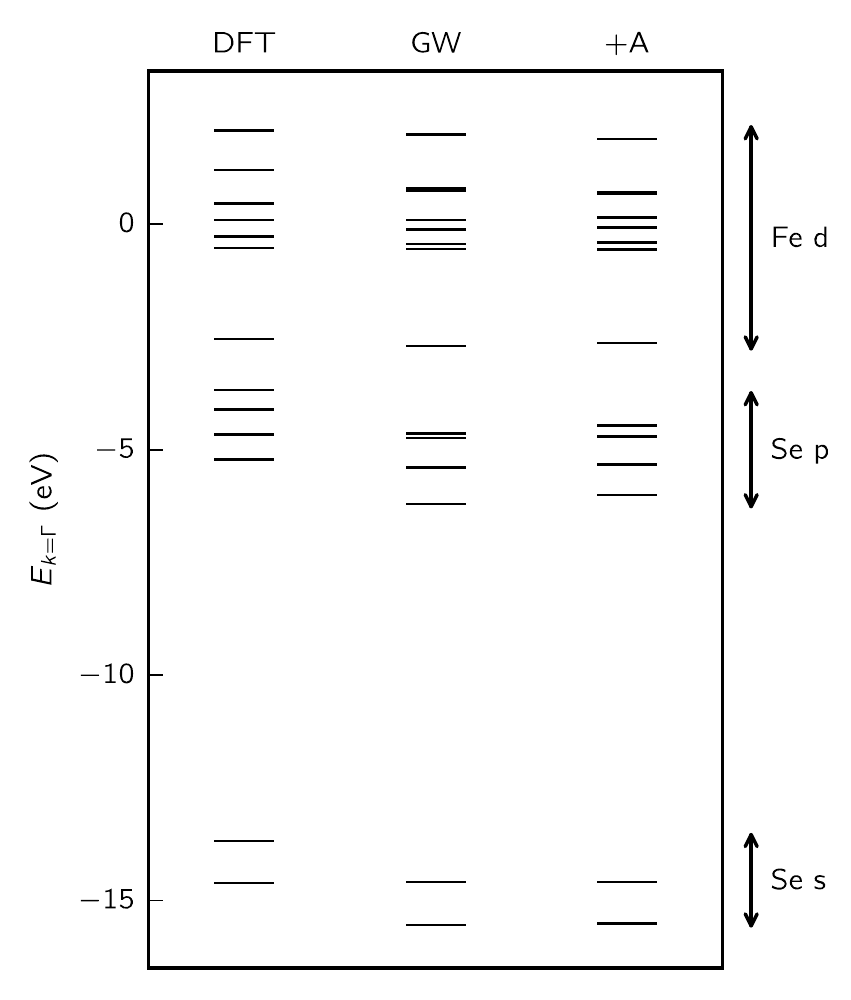}
  \caption{\label{fig:energylevels} Energy levels for occupied states at the $\Gamma$ point in monolayer FeSe as calculated at the DFT level, \GoWo{} level with the frequency-dependence in the screening captured within the HL-GPP model, and at the +A level with a static, semilocal approximation to the self energy.}
\end{figure}

In order to describe the electronic structure of FeSe, it is necessary to go beyond DFT and describe accurately the electron self energy using methods such as GW or DMFT. However, such methods tend to be expensive computationally. Next, we discuss the use of a local, static empirical potential to approximate the self energy (GGA+A). In this approach, we replace the exchange correlation potential $V_{GGA}(\mathbf{r})$ within the Kohn-Sham DFT with
\begin{equation}
V_{\rm GGA}(\mathbf{r}) + A\sum_i f(|\mathbf{r}-\mathbf{r}_i|),
\end{equation}
where $f(|\mathbf{r}-\mathbf{r}_i|)$ describes a repulsive potential centered around the position of each Fe atom ($\mathbf{r}_i$) and $A$ is an empirical fitting parameter. Previously, Ref.~\onlinecite{coh2015} showed that such an empirical correction, when fit to the experimentally known $M$ pocket width, greatly enhances the electron-phonon interaction in an FeSe monolayer. While these results were intriguing, it remained unclear how such a potential might affect the electronic structure for states far from the $M$-pocket, where the "A" parameter is fit, and for which no experimental data is available.

We fit the GGA+A expression in Eq. 1 to the $M$ pocket width from our \GoWo{} calculation with frequency-dependence of the screening described with the HL-GPP model. We used a potential of the form
\begin{equation}
    f(r) = e^{- r^2/a_0^2},
\end{equation}
where $a_0$ is the bohr radius and find a best fit with $A=0.25$~Ry. To mimic one-shot GW, in the results we present, we do not self-consistently update the GGA wavefunctions after adding +A. However, we find that self-consistency does not change the quality of the fit as long as $A$ is tuned. We find that with only a single parameter fit to reproduce the GW band structure near the $M$ pocket, GGA+A accurately reproduces the $GW$ energies at $\Gamma$ for the Fe $3d$ states as well as the low-lying Se $4s$ and $4p$ states (Fig.~\ref{fig:energylevels}). The GGA+A band structure near $E_F$ at the $M$ and $\Gamma$ points is shown in Figs.~\ref{fig:bsM} and ~\ref{fig:bsGamma}, respectively. The Fermi surface at the GGA+A level is shown in Fig.~\ref{fig:fermi}. We find that GGA+A qualitatively reproduces the changes to the band structure and the Fermi surface at the GW level (though the GGA+A gives a slightly larger Fermi surface) and agrees quantitatively with the GW energies to within 50 meV. This suggests that this method is surprisingly powerful, requiring only a single fitting parameter to reproduce most features of the a GW calculation with a computationally less expensive DFT--like calculation.

The good agreement between GW and GGA+A suggests that the real part of the GW self energy can be approximated by a local, static potential for this system. To better understand the dynamical and non-local contributions to the self energy, we examine the one-shot GW self energy in the static limit (static-COHSEX approximation). In the static-COHSEX approximation, the occupied band width at $M$ dramatically increases to 0.63~eV. The energy gap $\delta_M$ closes, and the lower $M$ point bands (labelled 3 and 4) cross the upper bands (labelled 1 and 2) so that the band maximum of band 3 and 4 is 0.21~eV higher than the bottom of the $M$ pocket. The electron pocket at $\Gamma$ disappears, as the entire band is pushed below the Fermi level.

In the dynamical GW calculation, the renormalization constant is
\begin{equation}
    Z_{n\textbf{k}}=(1-\frac{\partial\Sigma_{n\textbf{k}}(E)}{\partial E}|_{E=E^{\mathrm{QP}}_{n\textbf{k}}})^{-1},
\end{equation}
where ($n\textbf{k}$) are the band and wavevector indices, respectively, $\Sigma$ is the GW self energy and $E^{\mathrm{QP}}$ is the quasiparticle energy. $Z_{n\textbf{k}}$ gives the weight of the quasiparticle peak in the spectral function. For monolayer FeSe, $Z_{n\textbf{k}}$ is between 0.77 and 0.78 for all bands within 1 eV of the Fermi level at all k points in the Brillouin zone. Given the significant deviation of $Z_{n\textbf{k}}$ from 1, the large difference between the GW and static-COHSEX results is not surprising, but this raises the question of why the static GGA+A potential is so successful at reproducing the GW quasiparticle band structure, when the static limit of the GW approximation itself leads to very different results. The GW self energy can be written in terms of a Coulomb-hole term, $\Sigma^{\mathrm{COH}}$, and a screened-exchange term, $\Sigma^{\mathrm{SEX}}$. In the static-COHSEX approximation, $\Sigma^{\mathrm{COH}}$ can be written as a local potential, and the non-local contribution to $\Sigma^{\mathrm{SEX}}$ is generally small \cite{hybertsen86}. Thus, we might expect a tunable local potential to be able to approximate the static-COHSEX self energy.

To account for the dynamical effects, one must then analyze the source of the error in the static-COHSEX approximation, which comes from the assumption of an adiabatic accumulation of the Coulomb hole in the screened Coulomb interaction ~\cite{hedin1965,hedin1970,hybertsen86}. Numerically, Kang and Hybertsen have found that this error manifests in a different wavevector dependence between GW and static-COHSEX of $\Sigma^{\mathrm{COH}}$ and can be corrected by introducing a static scaling function in the Coulomb-hole term in the static-COHSEX approximation~\cite{kang2010}.
For the case of FeSe, we find that the difference between the static-COHSEX and GW self energies manifests primarily as a smooth wavevector-dependent shift in the magnitude of the self energy. In the vicinity of the Fermi energy, this shift is nearly uniform and thus easily captured by a tunable local potential of the form of $f(|\mathbf{r}-\mathbf{r}_i|)$ used in GGA+A.

\begin{figure*}[tp]
  \includegraphics{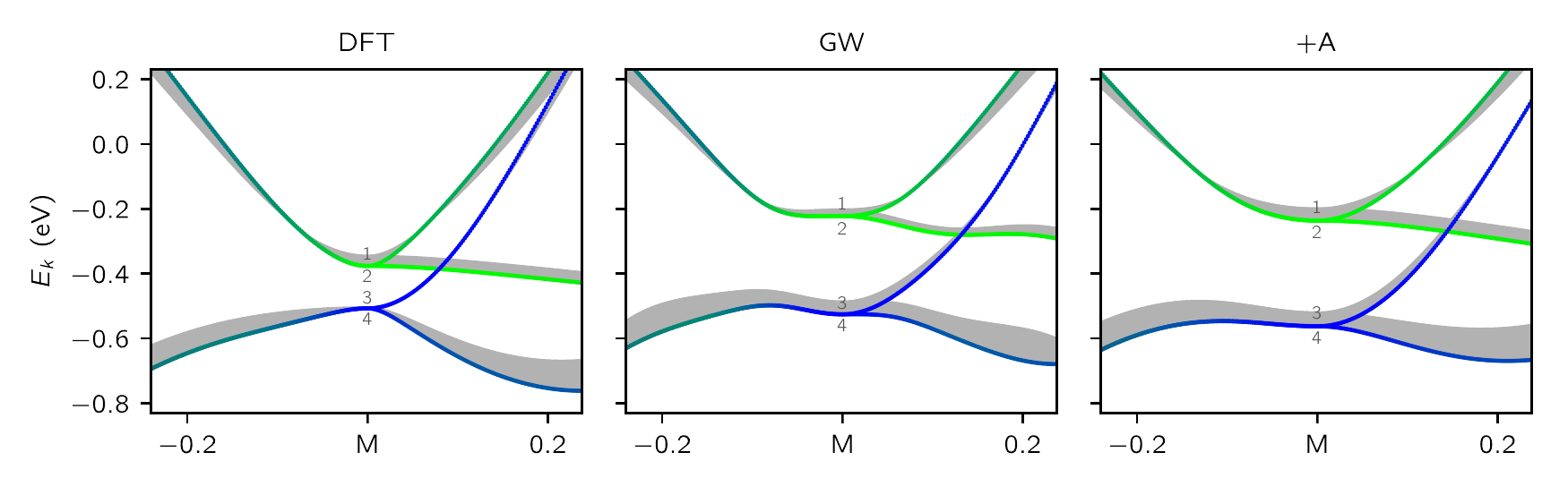}
  \caption{\label{fig:defpot} Colored lines are showing the band structure of the electron pocket near the $M$ point, exactly as in Fig.~\ref{fig:bsM}. The grayed areas show the evolution of the bands if the selenium height in FeSe is increased by 0.15~Bohr.}
\end{figure*}

In addition to the energy levels, we also examine the change in the band structure as the Se height is changed. The gray lines in Fig.~\ref{fig:defpot} show how the band structure energy near the $M$ point changes as the Se height is increased by 0.15~Bohr. The deformation potential for bands 1 and 2 (green) at the $M$-point is similar in magnitude for GGA-PBE, \GoWo{} with frequency-dependence in the dielectric screening at the HL-GPP level, and GGA+A (it is 35~meV, 24~meV, and 44~meV respectively). It is, however, very different for bands 3 and 4 (blue).  In DFT the change in the bands 3 and 4 with Se height displacement is 6~meV, while it is 43~meV and 46~meV in GW and GGA+A.

\section{Summary}

We present first-principles calculations of the electronic structure of monolayer FeSe at the GW level. We find that compared to DFT-GGA, GW increases the effective mass at the $M$ point, resulting in improved agreement with experiment.  Moreover, we show that the GW results for the quasiparticle band structure and deformation potentials can be reproduced to good accuracy at the DFT level with a semi-empirical correction involving only a single parameter, suggesting that such a correction, when parameterized by experiment or smaller-scale calculations at higher levels of theory, can be justifiably used to approximate the self energy correction to the band structure at greatly reduced computational cost.

\section{Acknowledgments}

This research was supported by the Theory of Materials Program at the Lawrence Berkeley National Lab through the Office of Basic Energy Sciences, U.S. Department of Energy under Contract No. DE-AC02-05CH11231, which provided the GW calculations and analyses. S.C. was supported by the National Science Foundation (NSF) under grant number DMR-1848074, which provided the GGA+A calculations. Advanced codes for the GW calculation were provided by the Center for the Computational Study of Excited State Phenomena (C2SEPEM), which is funded by the U.S. Department of Energy, Office of Science, Basic Energy Sciences, Materials Sciences and Engineering Division under Contract No. DE-AC02-05CH11231, as part of the Computational Materials Sciences Program. The National Energy Research Scientific Computing Center (NERSC), a DOE Office of Science User Facility supported by the Office of Science of the U.S. Department of Energy under Contract No. DE-AC02-05CH11231, provided computational resources for the GW calculations, and the Extreme Science and Engineering Discovery Environment (XSEDE), which is supported by National Science Foundation grant no. 787 ACI-1053575, provided computational resources for the DFT calculation.

\bibliography{fese_gw}

\end{document}